  \documentclass{aa}
\def\delv {\hbox{$\Delta v_{1/2}$}}

\def\TMB {$T_{\rm mb}$}

\def\H0 {$H_{\rm o}$}

\def\solmass {\hbox{M$_{\odot}$}}
\def\solum {\hbox{L$_{\odot}$}}

\def\kms {\hbox{${\rm km\,s}^{-1}$}}

\def\kmsmpc {\hbox{${\rm km\,s}^{-1}\,{\rm Mpc}^{-1}$}} 

\def\WAT {\hbox{${\rm H}_2{\rm O}$}}              
\def\CH3C2H {\hbox{${\rm CH}_3{\rm C}_2{\rm H}$}} 

\def \la{\mathrel{\mathchoice   {\vcenter{\offinterlineskip\halign{\hfil
$\displaystyle##$\hfil\cr<\cr\sim\cr}}}
{\vcenter{\offinterlineskip\halign{\hfil$\textstyle##$\hfil\cr
<\cr\sim\cr}}}
{\vcenter{\offinterlineskip\halign{\hfil$\scriptstyle##$\hfil\cr
<\cr\sim\cr}}}
{\vcenter{\offinterlineskip\halign{\hfil$\scriptscriptstyle##$\hfil\cr
<\cr\sim\cr}}}}}
\def \ga{\mathrel{\mathchoice   {\vcenter{\offinterlineskip\halign{\hfil
$\displaystyle##$\hfil\cr>\cr\sim\cr}}}
{\vcenter{\offinterlineskip\halign{\hfil$\textstyle##$\hfil\cr
>\cr\sim\cr}}}
{\vcenter{\offinterlineskip\halign{\hfil$\scriptstyle##$\hfil\cr
>\cr\sim\cr}}}
{\vcenter{\offinterlineskip\halign{\hfil$\scriptscriptstyle##$\hfil\cr
>\cr\sim\cr}}}}}

\begin{document}
\thesaurus{ 03          
           (02.13.3,    
            09.13.2,    
            11.01.2,    
            11.05.1,    
            11.09.4,    
            13.19.1)}   
\title{H$_2$O megamaser emission from FR\,I radio galaxies}
\author{C.~Henkel\inst{1,2}, Y.P.~Wang\inst{1,3}, H.~Falcke\inst{1,4}, 
A.S. Wilson\inst{4,5}, J.A. Braatz\inst{6}}
\offprints{C. Henkel, MPIfR, Auf dem H{\"u}gel 69, D-53121 Bonn, Germany
(p220hen@mpifr-bonn.mpg.de)}
\institute{
  Max-Planck-Institut f{\"u}r Radioastronomie,
  Auf dem H{\"u}gel 69, D-53121 Bonn, Germany
 \and
  European Southern Observatory, Casilla 19001, Santiago 19, Chile
 \and
  Purple Mountain Observatory, Academia Sinica, Nanjing 210008, China
 \and
  Dept. of Astronomy, University of Maryland, College Park, MD 20742, USA
 \and
  Space Telescope Science Institute, 3700 San Martin Drive, Baltimore,
  MD 21218, USA
 \and
  Harvard-Smithsonian Center for Astrophysics, 60 Garden St., Cambridge, MA
  21218, USA}
\titlerunning{H$_2$O megamasers in FR\,I radio galaxies}
\authorrunning{C.~Henkel et al.}
\date{Received 7 January 1998; accepted 7 April 1998}
\maketitle
\begin{abstract}

A systematic search for 22\,GHz \WAT \ megamaser emission is reported for 50
nearby ($z$ $\la $ 0.15) FR\,I galaxies. No detection was obtained, implying
that ultraluminous \WAT \ masers ($L_{\rm H_2O}$ $>$ 10$^{3}$\,\solum ) must
be rare in early-type galaxies with FR\,I radio morphology. Despite higher
radio core luminosities the detection rate for our sample is lower than
in similar surveys of late-type Seyfert galaxies. This puzzling difference
between Seyferts and low-power radio galaxies could be explained in several
ways: a) the maser emission is saturated and therefore independent of the
radio core luminosity, b) the masers are unsaturated and originate in a thin
circumnuclear gas disk, so the `seed' radio continuum would come from the 
far jet which is relativistically dimmed or c) the amount, kinematics, or
the distribution of the molecular gas in the nuclei of Seyferts and radio
galaxies is different. Further studies of maser properties may provide clues
to the differences between radio-loud and radio-quiet AGN. 


\end{abstract}

\section{Introduction} 

Systematic radio searches for \WAT \ emission among $\sim $ 600 galaxies led to
the detection of 17 `megamasers' and one `gigamaser' in active galactic nuclei
of the Seyfert 2 and LINER class (Braatz et al. 1994, 1996b, 1997; Koekemoer et
al. 1995; Greenhill et al. 1997; Hagiwara et al. 1997). Typical isotropic
luminosities are $L_{\rm H_2O}$ $\sim$ 100\,\solum; for the strongest
source $L_{\rm H_2O}$ $\sim$ 6100\,\solum\ is reached. Megamaser sources with
sufficiently intense \WAT \ emission allow high resolution interferometric
studies that provide interesting insights into active nuclear regions. This
includes the discovery of a Keplerian disk in the LINER NGC\,4258 (e.g.
Greenhill et al. 1995a,b; Miyoshi et al. 1995): Rotational velocity ($V_{\rm
rot} \sim $ 1000\,\kms ) and galactocentric radius ($R_{\rm GC} \sim $
0.15\,pc) require a nuclear mass concentration of density
$\ga$10$^{9}$\,\solmass \,pc$^{-3}$ that is consistent with a supermassive
black hole of a few 10$^{7}$\,\solmass \ (Maoz 1995).

Almost all extragalactic sources so far studied in \WAT \ are spirals. Few
early-type galaxies have been observed. Here we present a survey of FR\,I
(Fanaroff \& Riley 1974) radio sources including 50 low redshift ($z$ $<$
0.15) objects with $\delta $ $>$ --30$^{\circ }$ from the Zirbel \& Baum
(1995; hereafter ZB\,95) list of radio galaxies. Although the list is far from
representing a complete sample, it provides a large number of candidate
sources representative of this class of radio galaxy. The data were taken to
make a statistical comparison between Seyfert (i.e. radio quiet spiral) and
FR\,I (i.e. radio loud elliptical) galaxies w.r.t. \WAT\ detection rates, \WAT\
lineshapes, and nuclear cloud morphology. Strong differences would hint at
different types of gas distributions (e.g. tori) in the very central parts of
active galaxies (see Falcke et al. 1995).

\section{Observations and results} 

Observations in the 6$_{16} - 5_{23}$ transition of \WAT \ (rest frequency:
22.23508\,GHz) were obtained with the 100-m telescope of the MPIfR at
Effelsberg in December 1996 and March and April 1997. The beam width of the
telescope was 40$''$. A K-band maser receiver was used in conjunction with a
1024 channel, three level autocorrelator. In most cases the bandwidth of the
spectrometer was 50\,MHz, corresponding to $\sim $ 680\,\kms \ at the rest
frequency and providing a channel spacing of 0.66\,\kms \ and a velocity
resolution of 0.79\,\kms . Four sources were observed with a bandwidth of
25\,MHz. All observations were made in a position switching mode. Due to
superb weather in Dec. 1996, the system temperatures, including atmospheric
contributions, were 110 - 160\,K on a main beam temperature (\TMB ) scale.
During 1997, receiver temperatures were 150 -- 500\,K, depending on weather
conditions and elevation of the source.

Flux calibration was obtained by frequent measurements of NGC\,7027 and
3C\,286, which were assumed to have 22\,GHz flux densities of 5.86 and
2.55\,Jy, respectively (Baars et al. 1977; Ott et al. 1994). Variations of
these fluxes and of the receiver's calibration signal as a function of
frequency were taken into account; the data were also corrected for gain
variations of the telescope as a function of elevation. First to third order
polynomials were used to fit the baseline. Observational results are displayed
in Table 1. None of the observed 50 sources was detected. Single channel
1$\sigma $ noise limits, of order 100\,mJy in most cases, are given in
column\,7.

\scriptsize
\begin{table*}
\caption[]{\label{sourcelist} Observed sources$^{a)}$}
\begin{flushleft}
\begin{tabular}{lcrrccrrrrc}
\hline 
\multicolumn{11}{c}{\ } \\
\multicolumn{2}{c}{Source} &
\multicolumn{1}{c}{$\alpha_{1950}$} &
\multicolumn{1}{c}{$\delta_{1950}$} & 
\multicolumn{1}{c}{$P_{\rm core}^{\rm b)}$} &
\multicolumn{1}{c}{$T_{\rm int}^{\rm c)}$} &
\multicolumn{2}{c}{1\,$\sigma $ luminosity$^{\rm d)}$} &
\multicolumn{1}{c}{$V_{\rm min}^{\rm e)}$} &
\multicolumn{1}{c}{$V_{\rm max}^{\rm e)}$} &
\multicolumn{1}{c}{${\rm Epoch}^{\rm f)}$} \\
\multicolumn{5}{c}{ } &
\multicolumn{1}{c}{(min.)} &
\multicolumn{1}{c}{(mJy)} &
\multicolumn{1}{c}{(\solum )} &
\multicolumn{1}{c}{\kms } &
\multicolumn{1}{c}{\kms } &
\multicolumn{1}{c}{ } \\
\multicolumn{11}{c}{\ } \\
\hline
                &           &            &            &    &     &      & 
                                                      &      \\
3C\,29          &           & 00 55 01.6 & --01 39 44 & 23.58 & 31 &  47 & 
                                                 22.9 & 13200 &13700 & 3 \\
3C\,31          &           & 01 04 39.2 &  +32 08 44 & 22.04 & 12 &  34 &  
                                                  2.4 &  4800 & 5400 & 1 \\
PKS0115--261    &           & 01 15 52.8 & --26 07 35 & 23.35 & 25 & 123 & 
                                                 83.9 & 15600 &16200 & 3 \\
3C\,66B         &           & 02 20 01.7 &  +42 45 55 & 23.56 & 25 &  23 &  
                                                  2.7 &  6300 & 6900 & 1 \\
PKS\,0247--207  &           & 02 47 17.4 & --20 42 57 & 23.45 & 18 & 246 &
                                                452.2 & 25800 &26400 & 3 \\
3C\,75          &           & 02 55 03.2 &  +05 50 59 & 22.96 & 18 &  41 &  
                                                  5.3 & 6620 & 7220 & 3 \\
3C\,76.1        &           & 03 00 27.3 &  +16 14 33 & 22.66 & 25 &  33 &  
                                                  8.5 & 9550 & 9950 & 1 \\
3C\,78          &           & 03 05 49.1 &  +03 55 13 & 24.48 & 50 &  27 &  
                                                  5.5 & 8450 & 8950 & 1 \\
3C\,83.1B       &           & 03 14 56.8 &  +41 40 33 &       & 37 &  19 & 
                                                  3.1 & 7650 & 7950 & 1 \\
3C\,84          & NGC\,1275 & 03 16 29.6 &  +41 19 52 &       & 12 &  46 &  
                                                  3.5 & 5100 & 5450 & 1 \\
3C\,89          &           & 03 31 42.4 & --01 22 21 & 24.53 & 25 &  85 &
                                                396.5 &41350 &41800 & 3 \\
3C\,120         &           & 04 30 31.6 &  +05 15 00 & 25.19 & 25 &  40 & 
                                                 10.8 & 9600 &10300 & 2 \\
                &           &            &            &       & 56 &  31 &
                                                  8.2 & 9600 &10200 & 3 \\
PKS\,0449--175  &           & 04 49 07.0 & --17 35 12 & 21.58 & 12 & 121 & 
                                                 28.2 & 9000 & 9600 & 2 \\
PKS\,0545--199  &           & 05 45 45.1 & --19 59 03 & 23.23 & 12 & 189 &
                                                128.9 &15600 &16200 & 2 \\
PKS\,0634--205  &           & 06 34 22.3 & --20 32 14 & 22.74 & 12 & 161 &
                                                118.3 &16200 &16800 & 2 \\
                &           &            &            &       & 12 & 182 &
                                                133.7 &16200 &16800 & 3 \\
B2 0722+30      &           & 07 22 27.6 &  +30 03 13 & 22.90 & 25 &  22 &
                                                  1.9 & 5550 & 5850 & 1 \\
B2 0800+24      &           & 08 00 16.3 &  +24 49 02 & 22.38 & 12 &  24 & 
                                                 11.6 & 13250 &13550 & 1 \\
3C\,218         &           & 09 15 40.8 & --11 53 10 & 24.51 & 12 & 159 &
                                                163.2 &19200 &19800 & 2 \\
B2 0915+32      &           & 09 15 57.4 &  +32 04 20 & 23.12 & 12 &  30 & 
                                                 27.9 &18250 &18900 & 1 \\
B2 1108+27      &           & 11 08 44.1 &  +27 13 48 &       &  6 &  83 & 
                                                 22.0 & 9600 &10200 & 1 \\
B2 1122+39      & NGC\,3665 & 11 22 01.3 &  +39 02 17 & 21.06 & 12 &  44 &  
                                                  0.5 & 1700 & 2300 & 1 \\
3C\,264         &           & 11 42 29.1 &  +19 53 05 & 23.58 & 12 &  40 &
                                                  4.3 & 6000 & 6600 & 2 \\
B2 1144+35      &           & 11 44 12.6 &  +35 45 09 & 24.62 & 12 &  50 & 
                                                 48.2 &18550 &19250 & 1 \\
3C\,270         & NGC\,4261 & 12 16 49.9 &  +06 06 09 & 22.68 & 56 &  16 &
                                                  0.2 & 1950 & 2500 & 1 \\
3C\,272.1       &           & 12 22 31.5 &  +13 09 50 & 21.92 & 18 &  27 &
                                                  0.1 &  600 & 1200 & 2 \\
3C\,274.0       & NGC\,4486 & 12 28 17.8 &  +12 39 58 & 23.37 & 18 &  36 &
                                                  0.1 &  900 & 1500 & 2 \\
3C\,278         &           & 12 51 59.6 & --12 17 08 & 22.87 & 18 &  40 &  
                                                  2.0 & 3900 & 4800 & 2 \\
B2 1254+27      &           & 12 54 02.4 &  +27 15 27 &       & 25 &  21 &  
                                                  2.4 & 6150 & 6750 & 1 \\
                &           &            &            &       & 12 &  42 &
                                                  6.2 & 7100 & 7700 & 2 \\
B2 1317+33      &           & 13 17 58.7 &  +33 24 24 &       & 37 &  16 &
                                                  5.6 &11050 &11700 & 1 \\
B2 1322+36      &           & 13 22 35.3 &  +36 38 18 & 23.29 & 31 &  26 &  
                                                  2.0 & 5050 & 5650 & 2 \\
B2 1346+26      &           & 13 46 34.2 &  +26 50 25 & 23.95 & 12 &  46 & 
                                                 44.2 &18650 &19100 & 1 \\
3C\,296.0       &           & 14 14 26.4 &  +11 02 19 & 23.27 & 12 &  45 &
                                                  6.3 & 6900 & 7500 & 2 \\
B2 1422+26      &           & 14 22 26.5 &  +26 51 02 & 23.16 & 12 &  33 & 
                                                 11.0 &10800 &11400 & 1 \\
3C\,315         &           & 15 11 30.8 &  +26 18 40 &       &  6 &  23 & 
                                                 65.6 &32150 &32850 & 1 \\
3C\,317         &           & 15 14 17.1 &  +07 12 15 & 24.50 & 12 &  34 &
                                                  9.6 & 9950 &10550 & 1 \\
B2 1525+29      &           & 15 25 39.6 &  +29 05 28 & 22.66 & 12 &  59 & 
                                                 61.2 &19250 &19950 & 1 \\
B2 1553+24      &           & 15 53 56.2 &  +24 35 33 &       & 12 &  26 & 
                                                 11.5 &12450 &13100 & 1 \\
NGC\,6047       & NGC\,6047 & 16 02 54.0 &  +17 51 51 &       & 12 &  31 &  
                                                  7.4 & 9150 & 9650 & 1 \\
                &           &            &            &       & 18 &  77 & 
                                                 18.6 & 9150 & 9750 & 2 \\
B2 1610+29      &           & 16 10 35.6 &  +29 36 41 &       & 12 &  56 & 
                                                 13.2 & 9100 & 9600 & 1 \\
B2 1621+38      &           & 16 21 16.9 &  +38 02 16 & 23.31 & 12 &  36 &  
                                                  8.4 & 9000 & 9600 & 1 \\
3C\,338         &           & 16 26 55.7 &  +39 39 40 & 23.61 & 18 &  72 & 
                                                 15.7 & 8700 & 9300 & 2 \\
3C\,386         &           & 18 36 12.0 &  +17 09 09 &       & 12 &  35 &  
                                                  2.5 & 4800 & 5400 & 1 \\
B2 1855+37      &           & 18 55 54.3 &  +37 56 26 &       & 12 &  20 & 
                                                 14.6 & 16200 &16700 & 1 \\  
3C\,424         &           & 20 45 44.4 &  +06 50 12 &       &  6 & 456 &
                                                1988.7&39900 &40500 & 2 \\
                &           &            &            &       & 18 & 512 &
                                                2005.7&37800 &38400 & 3 \\
PKS\,2104--25   &           & 21 04 30.0 & --25 37 54 & 23.25 & 37 & 213 & 
                                                 69.2 &10770 &11170 & 3 \\
                &           &            &            &    &     &      & 
                                                      &      \\
\hline
\end{tabular}
\end{flushleft}
\end{table*}

\begin{table*}
{\bf Table \thetable{}.} Continued
\begin{flushleft}
\begin{tabular}{lcrrccrrrrc}
\hline 
\multicolumn{11}{c}{\ } \\
\multicolumn{2}{c}{Source} &
\multicolumn{1}{c}{$\alpha_{1950}$} &
\multicolumn{1}{c}{$\delta_{1950}$} & 
\multicolumn{1}{c}{$P_{\rm core}^{\rm b)}$} &
\multicolumn{1}{c}{$T_{\rm int}^{\rm c)}$} &
\multicolumn{2}{c}{1\,$\sigma $ luminosity$^{\rm d)}$} &
\multicolumn{1}{c}{$V_{\rm min}^{\rm e)}$} &
\multicolumn{1}{c}{$V_{\rm max}^{\rm e)}$} &
\multicolumn{1}{c}{${\rm Epoch}^{\rm f)}$} \\
\multicolumn{5}{c}{ } &
\multicolumn{1}{c}{(min.)} &
\multicolumn{1}{c}{(mJy)} &
\multicolumn{1}{c}{(\solum )} &
\multicolumn{1}{c}{\kms } &
\multicolumn{1}{c}{\kms } &
\multicolumn{1}{c}{ } \\
\multicolumn{11}{c}{\ } \\
\hline
                &           &            &            &    &     &      & 
                                                      &      \\
B2 2116+26      & NGC\,7052 & 21 16 20.8 &  +26 14 08 & 22.73 & 12 &  33 &  
                                                  2.1 & 4600 & 5200 & 1 \\
3C\,449         &           & 22 29 07.6 &  +39 06 03 & 22.72 & 12 &  43 & 
                                                  3.2 & 4950 & 5550 & 1 \\
4C+11.71        &           & 22 47 25.1 &  +11 20 36 &       & 12 &  34 &  
                                                  5.6 & 7500 & 8100 & 1 \\
PKS\,2322--123  &           & 23 22 43.7 & --12 23 57 & 23.31 & 37 & 144 &
                                                236.1 &24530 & 24770 & 3 \\
3C\,465         &           & 23 35 59.0 &  +26 45 16 & 24.02 & 12 &  33 &
                                                  6.7 & 8400 & 9000 & 1 \\
                &           &            &            &    &     &      & 
                                                      &      \\
\hline
\end{tabular}
\ \ \ \\
\ \ \ \\
a) For radio properties and optical emission line parameters, see ZB\,95. \\
b) Logarithm of the 5\,GHz radio core power in units of W/Hz (see  ZB\,95) \\
c) $T_{\rm int}$ (column 6) includes Effelsberg 22\,GHz (\WAT) on- and 
   off-source integration time. 
   \\
d) Root mean square (rms) flux density in one spectral channel (nominal
   width: 0.66\,\kms ; velocity resolution: 0.79\,\kms ).
   Quoted (isotropic) \WAT\ luminosities are [$L$/\solum ] = 0.023 [rms/Jy] 
   [\delv /\kms ] [D/Mpc]$^{2}$ for a 0.66\,\kms \ wide channel; 
   $H_{\rm o}$ = 75\,\kmsmpc . Luminosities of detected sources are 
   20 $\leq$ $L_{\rm H_2O}$/\solum\ $\leq$ 6100 (BWH\,96). \\
e) Approximate optical velocity range searched. V is related 
   to the observed frequency by $\nu _{\rm observed}$ = $\nu_{\rm rest}$ 
   (1 + $V_{\rm opt}$/c)$^{-1}$. \\
f) 1: Dec. 22 -- 24, 1996; 2: Mar 24 -- 25, 1997; 3: Apr. 7 -- 9, 1997 \\
\end{flushleft}
\end{table*}
\normalsize

\section{Early-type radio galaxies versus spirals}

\subsection{FR\,I, FR\,II, and spiral samples}

The FR\,I sources displayed in Table 1 include 75\% of all the ZB\,95 FR\,I
galaxies and 89\% of the ZB\,95 FR\,I sources with declination $\delta $ $>$
--30$^{\circ}$ and redshift $z$ $<$ 0.15. Hubble types from the NASA/IPAC
Extragalactic Database (NED) are consistent with an early-type nature. Data
from previous surveys include non-detections of \WAT\ emission from the ZB\,95
FR\,I sources 3C\,84, B2 0915+32, B2 1122+39, 3C\,274.0, B2 1254+27, B2
1525+29, B2 1553+24, and B2 2116+26 (Braatz et al. 1996b, hereafter BWH\,96).
All these galaxies are also observed by us. 22\,GHz \WAT \ upper limits for
FR\,II galaxies of the ZB\,95 sample were reported from 3C\,390.3 and Cyg\,A
(3C\,405) (BWH\,96). The sample of ZB\,95 FR\,II upper limits is therefore too
small for a statistical evaluation.

The Braatz et al. (1997; hereafter BWH\,97) spiral sample contains 193 observed
galaxies (0 $\leq $ $T$ $\leq $ 9 according to the de Vaucouleurs et al.
(1991) notation) and 10 \WAT \ detections; including the four spirals that
would have been part of the sample if they had not been detected before (see 
Braatz et al. 1994), the detection rate is 7.1$\pm $1.6\% (the error being the
standard deviation from the Bernoulli theorem). For the distance limited
sample (c$z$ $\la$ 7000\,\kms ), the corresponding values are 176, 14, and
8.0$\pm $1.8\%. Our sample of nearby (c$z$ $\la$ 7000\,\kms ) FR\,I objects
only comprises 15 sources. All of these were measured with high sensitivity
(1$\sigma $ $\la $ 5\,\solum \ for an individual channel), so that any
megamaser would have easily been seen. Assuming the same detection rate as for
the distance limited BWH\,97 sample leads to 1.2$\pm $0.3 expected detections
that do not deviate significantly from our null result.

To estimate detection probabilities for our entire sample of galaxies, we note
that the spiral BWH\,97 and the early-type FR\,I sample differ by the power and
distance of their radio cores. For the distance limited BWH\,97 sample, few
nuclear $\lambda $ = 1.3\,cm radio fluxes are known; at $\lambda $ = 6\,cm,
however, VLA A or B array data are available for 45 (26\%) sources. For the
corresponding distance limited FR\,I sample, 13 (87\%), and for the more
distant (c$z$ $\ga$ 7000\,\kms ) FR\,I galaxies, 24 (69\%) sources have known
$\lambda $ = 6\,cm core fluxes. The FR\,I core radio luminosities are larger
by almost one and more than two orders of magnitude than those of the spiral
sample. {\it Assuming unsaturated maser amplification of the radio core}, the
observational sensitivity for \WAT \ 22\,GHz radiation depends almost linearly
on nuclear radio fluxes (e.g. Reid \& Moran 1981). Assuming similar radio
source covering factors, {\it the overall sensitivity of our maser search is},
for both the nearby and the distant FR\,I sources, {\it higher than that for
the BWH\,97 spiral sample}. The sensitivity ratio is of order 10.

To estimate detection probabilities, some qualitative knowledge of the 
\WAT\ luminosity function is also needed. In their Table 1, BWH\,96 list
eight masers with 10--100\,\solum, seven with 100--1000\,\solum , and one with
6100\,\solum . We note that the BWH\,96 detection rate decreases with
increasing redshift (their Figs.\,2--4) which implies that some of the less
luminous ($L_{\rm H2O}$ $<$ 100\,\solum ) masers are not detected because of
sensitivity limits. We also note that the location of the gigamaser (Koekemoer
et al. 1995) is slightly outside the systematically studied c$z$ $\la$
7000\,\kms \ redshift range. Since sources in the gigamaser range are
definitely rare, the spatial density of \WAT \ masers of a given luminosity is
consistent with $n_{\rm H2O}$ $\propto $ $L_{\rm H2O}^{-\alpha}$, $\alpha$ $>$
0.

If differences in distance and Hubble type are ignored, detection rates for
our FR\,I sample should be identical to that of the spiral BWH\,97 sample (8.0
$\pm$ 1.6\%). With a `reasonable' luminosity function ($\alpha$ $\ga$ 0.5) and
accounting for the net sensitivity gain of our FR\,I versus the BWH\,97 spiral
survey in the case of unsaturated amplification of the radio core (radio power
versus distance squared), the detection rate for FR\,I galaxies becomes at
least five times higher than that for spirals with AGN. The expected FR\,I
detection rate then is $\ga$40\%.

\subsection{Interpretation}

In view of our unsuccessful search for \WAT \ masers in FR\,I galaxies, {\it
expected detection rates inferring unsaturated maser emission of the radio
core are far too large}. What is causing the discrepancy between expected and
measured values?

\begin{itemize}

\item Rotating disk scenario: If unsaturated maser amplification occurs in a
slightly inclined (i.e. not perfectly edge-on) circumnuclear thin disk, the core
radio flux is not amplified. Instead, the background arises from the receding
counter jet that, relativistically dimmed, may be too weak a source of 
`seed' photons. 

\item Saturated maser emission: 
{\it If} masers are saturated (see Greenhill et al. 1995a; Braatz et al.
1996a; Kaufman \& Neufeld 1996; Trotter et al. 1996; Herrnstein et al. 1997;
Claussen et al. 1998), we can relax the assumed correlation between radio core
and maser luminosity and obtain with the detection rate for the distance
limited spiral BWH\,97 sample and $\alpha$ $\ga$ 1 (see Sect.\,3.1) a total of
1.4$\pm $0.4 expected FR\,I galaxy detections. This is consistent with our
observational result.

\item Very broad line emission: The limited velocity coverage of our spectra,
$\Delta V$ $\sim$ 600\,\kms, leaves chances for further improvement. The
superposition of thousands of individual maser components, covering a velocity
range $>$1000\,\kms, or rapidly rotating tori with only the tangential parts
showing strong (highly red- and blue-shifted) maser emission are possible
scenarios consistent with an absence of detections. 

\item Lack of molecular gas: Towards a number of prominent radio galaxies, HI
absorption was detected that may be associated with the nuclear region (e.g.
van Gorkom et al. 1989; Conway \& Blanco 1995; Henkel \& Wiklind 1997). With
the notable exception of Cen A, however, nuclear CO has not been seen with
certainty. It remains open whether this is caused by amount and composition of
the nuclear gas or whether it is merely a consequence of observational
sensitivity limits (see Henkel \& Wiklind 1997 for a review).
 
\end{itemize}

\section{Conclusions}

Having searched in 50 nearby ($z$ $<$ 0.15) FR\,I galaxies for 22\,GHz \WAT \
emission, our main conclusions are as follows:

(1) No new megamaser has been deteted.

(2) Our detection rate is much smaller than that for spiral galaxies with AGN
if unsaturated maser emission, amplifying the observed nuclear continuum, is
assumed.

(3) Our negative result can be explained in terms of maser saturation (i.e. a
large optical depth in the maser line), in terms of unfavorable geometries
minimizing the number of available 22\,GHz seed photons (e.g. a thin,
circumnuclear gas disk which projects in front of the more distant,
relativistically dimmed jet), by different kinematics of the molecular material
(e.g. leading to very broad lines), or by a general lack of warm dense
molecular gas in the nuclear regions of FR\,I galaxies.

At present we cannot decide which of the possibilities given here is
more likely to explain our low detection rate. However, if further
searches in larger samples of low-power radio galaxies continue to
yield negative detection rates, an explanation in terms of saturated
maser emission alone may no longer be viable and could point to an
intrinsic difference in the nuclear properties of Seyferts and
low-power radio galaxies. Such a result could then be corroborated by
further and more sensitive observations of molecular lines and VLBI
observations of the continuum sources in Seyferts and low-power radio
galaxies.

\acknowledgements{This research was supported by NATO grant SA.5-2-05
(GRG.960086) 318/96. HF was supported in part by the DFG, grant Fa
358/1-1\&2.}

\end{document}